\documentclass[twocolumn]{IEEEtranUSMA}
\usepackage[dvips]{graphicx}
\usepackage{fancyheadings}
\usepackage{amsmath}
\usepackage[latin1]{inputenc}
\usepackage{url}

\newcommand{\com}[1]{\texttt{#1}} 
\newcommand{\buzz}[1]{\textit{#1}} 

\begin{document}


\title{NoSEBrEaK -- Attacking Honeynets} 
\author{Maximillian Dornseif
\thanks{dornseif@informatik.rwth-aachen.de -- Laboratory for Dependable Distributed Systems, 
RWTH Aachen University.} 
Thorsten Holz
\thanks{holz@i4.informatik.rwth-aachen.de -- Laboratory for Dependable
Distributed Systems, RWTH Aachen University.} 
Christian N. Klein 
\thanks{kleinc@cs.bonn.edu -- University of Bonn.}
}  



\pagestyle{fancy}
\setlength{\headrulewidth}{0pt}
\lhead{}

\lfoot{}
\cfoot{}
\rfoot{\thepage}

\lhead{}
\rhead{}

\maketitle
\thispagestyle{fancy}


\begin{abstract}
  It is usually assumed that Honeynets are hard to detect and that attempts to
  detect or disable them can be unconditionally monitored. We
  scrutinize this assumption and demonstrate a method how a host in a
  honeynet can be completely controlled by an attacker without
  any substantial logging taking place. 
\end{abstract}


\section{Introduction}
\label{sec:intro}

At the Laboratory for Dependable Distributed Systems at \mbox{RWTH} Aachen
University, Germany, we run a Linux based honeynet for gathering information
on security incidents.  The scientific method dictates that we must attack our
own assumptions vigorously to get meaningful results.  Under the code name
``NoSEBrEaK'' we attacked our original assumptions about undetectability and
monitorability of honeynets by turning the table and taking the view of an
attacker trying to control a honeypot.  In the following paper we present the
results of this red team approach.

Honeynet researchers want to ``learn the tools, tactics, and motives of the
blackhat community and share the lessons learned''\cite{honeynet:homepage}.
Therefore honeynets must provide a way to capture the activities of an
intruder in order to provide some information about his actions and the tools
he used. The traditional method used to capture keystrokes and other
activities on Unix Systems is to patch \com{/bin/sh} -- the standard
interpreter -- in order to log every keystroke of an intruder to a special
file. Other possibilities are to redirect the output of \com{syslogd} to
another host on the network or to record and analyze all network traffic with
\com{tcpdump} and other tools. But network monitoring attempts are doomed to
fail if the intruder utilizes encryption -- for example by using SSH or SSL to
connect to the honeynet -- to protect his communication channel. Then all
captured data is useless because the information can not be decrypted without
the appropriate key. 

Trojaned binaries on the honeynet that record all keystrokes by the attacker
can be circumvented if the attacker installs his own binaries, which is a
common way nowadays. Thus, a new way to capture the activities of an intruder
on an Linux system was developed: A kernel-based rootkit called \buzz{Sebek}
\cite{sebek:homepage} is able to record all data accessed by users via the
\com{read()} system call on the honeynet. It replaces the normal \com{read()}
system call with a new entry in the system call table pointing to its own
version of this system call. It is then able to record all data accessed via
\com{read()} \cite{Sebek:2003}. Because Sebek lives entirely in kernel-space
and has access to all data read, this rootkit is able to access most
communication unencrypted and it can for example log SSH-sessions, recover
files copied with SCP and record all passwords used by intruders. The recorded
data is sent via UDP to the Sebek server, the other part of Sebek's
client/server architecture. This transmission is done by modifying the kernel
in order to hide these packets such that an intruder can not see them. In
addition, all network counters and data structures have to be readapted in
order to make detecting these changes more difficult. Further information
about Sebek and its architecture can be found in \cite{Sebek:2003}.

In this paper we show that a qualified attacker -- in contrast to an
unsophisticated attacker or autonomous malware like worms -- has several ways
to detect and circumvent Sebek, the primary data capture tool used by honeynet
researchers to capture the attackers' activities on a honeynet.

The paper is outlined as follows: Section \ref{sec:related} gives an overview
of related work in the field of detection of honeynets. Several ways to
detect, disable and bypass Sebek, mainly implemented in our proof of concept
toolkit \buzz{Kebes}, are presented in section \ref{sec:kebes}. Directions of
further work are outlined in section \ref{sec:further} and we conclude this
paper with section \ref{sec:con}.

\section{Related work}
\label{sec:related}

Unfortunately, there is little scientific literature on the topic of honeynets
and none on detecting or circumventing data capture mechanisms in honeynets.
The only related work was published in two fake issues of Phrack
\cite{phrack:homepage}, a Hacker magazine that is famous for its articles in
the blackhat community.

In Phrack 62 \cite{corey03:_local_honey_ident} an article entitled ``Local
Honeypot Identification'' was published. It describes a method to disable
Sebek by just overwriting Sebek's \com{read()} system call in the system call
table with the original value and thus disabling Sebek. We could not verify
that this is easily achievable since it seems that the position of the
original system call can not easily obtained.  Other techniques to detect the
existence of Sebek on a host were also presented in the article. They include
detection of an outgoing connection limit used on most honeynets and the
existence of \buzz{snort\_inline} \cite{snortinline:homepage}, an intrusion
prevention systems that modifies packets that look harmful.

Another idea to defeat Sebek was published in issue 63 of the Phrack magazine
\cite{corey04:_advan_honey_pot_ident}: The article ``Advanced Honey Pot
Identification'' describes a way to search through memory and look for
characteristic symbols used by Sebek. The accompanying code is claimed to be
able to reconstruct several highly sensitive bits of Sebek data, including the
\buzz{magic number} and \buzz{port numbers} used by Sebek to identify and hide
its communication from others.  After publication of the article, multiple
changes in Sebek's code were released to counter this identification
techniques, including randomization of the symbol names used. 

Issue 61 of the Phrack magazine contained an article on detection of hidden
kernel modules in its ``Linenoise'' section \cite{madsys03:_findin}. The
article describes a brute force method for detecting hidden modules by looking
for what appears to be the key module structure. In our tests this method was
able to locate a hidden Sebek module.

There also is a tool named \buzz{chkrootkit} which ``is a tool to locally
check for signs of a rootkit'' \cite{chkrootkit:homepage}. While the
documentation claims that the tool is able to detect Sebek, this is only true
for very old versions of Sebek. Detecting is accomplished by searching
\com{/proc/ksyms} for the string ``sebek''.

\section{Several ways to detect and circumvent Sebek}
\label{sec:kebes}

In this section we present several ideas and ways to detect and circumvent
Sebek in order to prevent logging of user activities and thus effectively
preventing honeynet researches to collect information about intruders.

We first present ideas to detect Sebek, followed by sections dealing with
disabling of Sebek and detection of the honeywall. In the last section we
describe several ideas and ways to circumvent Sebek, in particular our
implementation \buzz{Kebes}, a tool to effectively bypass Sebek.

\subsection{Detecting Sebek}
\label{sub:detect}

Sebek is a kernel module, so it leaves several traces of its existence in
memory. We now describe ways to track down those traces, thus leading to the
detection of Sebek.

One of our first experiments originally was aimed at making logging by Sebek
unreliable. By accident we found an easy way to detect Sebek which can be used
even by unprivileged local users:

\begin{verbatim}
  user@honey:~ dd if=/dev/zero of=/dev/null bs=1
\end{verbatim}

This simple command results in backward running packet and byte counters of
the network interface, so even an unprivileged user was able to detect Sebek
by checking the output of \com{ifconfig}. After we pointed this out to Sebek's
maintainers, this issue was mitigated with version 2.1.7 of Sebek.

To understand why using \com{dd} should lead to unreliable logging by Sebek,
one has to take a closer look at the logging process: As explained above,
Sebek replaces the normal \com{sys\_read()} call in the system call table with
its own version and is thus able to record all data accessed via \com{read()}.
The logging data is then transfered in a covert way via UDP over the
network\cite{Sebek:2003}. Therefore, if we are reading a single byte via a
\com{read(1)}, Sebek has to transfer nearly 100 Bytes of data including all
the network headers over the network to the host doing the logging. So if we
are able to do a \com{read(1)} some tens of thousands of times per second,
this will lead to a congested network and finally to dropped packets. And we
are able to identify a congested network with the help of the \buzz{ping}
command.

This leads to another way to detect Sebek, which we call the \buzz{dd-attack},
since it can accomplished with a simple command like \com{dd}, as shown above.
To detect Sebek on a host, we first ping a local IP-address (e.g. the gateway)
in order to get an overview of the current performance of the network.  Then
\buzz{dd} is started in the background and we rerun the ping command. If Sebek
is installed on a host, this leads to an enormous increase in the average
round-trip time. In practical tests of this attack the average round-trip time
grew from 0.7 milliseconds to over 4800 milliseconds.

But the fix in Sebek which was introduced in version 2.1.7 leads to a new way
to detect Sebek: In an attempt to control the backward moving
transmit-counters, the counters are untouched and Sebek modifies the output
routines for \com{/proc/net/dev} to deduct the number of Sebek packages from
the total number of transmitted packages just before output. This enables us
to compare \com{dev->get\_stats()->tx\_bytes} with the values listed in
\com{/proc/net/dev} for every ethernet device on the system. Differences can
only be explained by the presence of Sebek.

Another way to detect Sebek is similar to the approach published in Phrack 63
\cite{corey04:_advan_honey_pot_ident}: Sebek is a kernel module and it has to
use some kernel memory in order to be able to store its variables like
\buzz{magic number} and \buzz{source} or \buzz{destination} port. Version
2.1.7 uses some randomization and other stealth techniques in order to prevent
easy ways to find these variables, but Sebek leaves some traces in memory that
can be found:

\begin{itemize}
  
	\item Since Sebek by default chooses a random number below
		$1000000000$ as the module name it is easy to create
		heuristics to detect the Sebek module purely by its name in
		the lists of modules generated by the methods outlined in
		\cite{madsys03:_findin}. A small obstacle in doing so is that
		there are legitimate modules with purely numeric names like
		\com{8390}, a driver for a generic Ethernet chipset.
         
	\item Helpful in the detection of Sebek symbols is the knowledge that
		all it's symbol names start with a letter, followed by a
		number with up to three digits.
         
	\item Also helpful is the fact that a conversion table to translate IP
		addresses into integers is left in memory. In addition to
		falsify \com{/proc/net/dev} output, Sebek has to keep more
		than 200 bytes of strings in memory, which can be located
		without great effort.

	\item The hidden Sebek module still has a complete module header
		structure in memory. This structure can be found with very
		little effort\cite{madsys03:_findin} and used to locate
		Sebek's code (e.g.its cleanup function) and variable storage. 

	\item While the placement of variables in Sebek's variable storage
		block is different every time Sebek is compiled, it is still
		possible to locate the variable storage block via the module
		header. That information can be used to successfully guess
		most variables like \buzz{magic number}, \buzz{source} and
		\buzz{destination} port with a few tries.

\end{itemize}

All these traces can be found by traversing memory and looking for appropriate
structures and thus leading to a way to detect hidden kernel modules, identify
the Sebek kernel module and extract secrets from the module.  We are able to
implement this approach and it leads to another way by which Sebek can be
detected on a host.

In \cite{corey03:_local_honey_ident}, a way to detect the changes in the
system call table for Solaris is proposed. We carried this idea forward and
implemented it: When Sebek installs its \com{read()} monitoring function, it
replaces the function pointer in the system call table at \com{NR\_read} with
a pointer to its own function, declared in the module. While on a system
without Sebek, the system call pointers are close to one another in memory,
this changes if Sebek is loaded into memory. The address of Sebek's function
does not lie near the address of \com{sys\_write}, the successor of
\com{sys\_read} in the system call table. Thus, if the addresses of
\com{sys\_read} and \com{sys\_write} lie far apart (e.g. more than 4096 byte),
we conclude that the system call table was modified by some rootkit.
Furthermore, there is another similar way to detect Sebek: After Sebek has
replaced the function pointer in the system call table, \com{sys\_read} points
inside the memory range of a module. This is a clear sign that the system
calls were changed and a rootkit is present. We wrote a program that searches
through memory and finds both of these anomalies. 

As an example, Figure \ref{fig:sys} shows the places of two system calls in
memory before and after Sebek is loaded.  One can clearly see that the system
call table was modified and something malicious is happening.

\begin{figure}[htb]
\begin{verbatim}
before:       __NR_read  = 0xc0132ecc
              __NR_write = 0xc0132fc8
              
after:        __NR_read  = 0xc884e748
              __NR_write = 0xc0132fc8
\end{verbatim}
\caption{Addresses of two system calls before and after Sebek is loaded.}
\label{fig:sys}
\end{figure}

\subsection{Disabling Sebek}
\label{subsec:disable}

After having shown several ways to detect Sebek, we now point out sundry
possibilities to disable it. One evident possibility to disable Sebek is
simply rebooting. Currently Sebek offers no possibility to be loaded
automatically on system startup, but this can easily be achieved by
customizing the shell scripts. But integrating the installation of Sebek into
the boot process would result in leaving a lot of evidence of Sebek's
existence and an avenue for removing this boot scripts and rebooting to
disable Sebek.

Another possibility is more complicated: Reconstruction of the original system
call table and thus effectively disabling Sebek. We were not able to reproduce
results from \cite{corey03:_local_honey_ident}: Since \com{sys\_read} is not
an exported kernel symbol, we can not simply obtain the address of the
original \com{sys\_read}. It might be feasible to find the memory address of
the original \com{sys\_read} call through parsing the kernel image on disk. An
alternative would be searching through the memory of the Sebek kernel module
to obtain the address of the original system call.

Probably the most elegant solution is asking Sebek to remove itself.  If we
are able to obtain the address of Sebek's \com{cleanup()} function we can call
this function which will result in Sebek restoring the original state of the
system call table and thus disabling further logging. Since we have already
shown that the key module structures can be reconstructed, we are able to
implement this approach to disable Sebek.

\subsection{Detecting the Honeywall}
\label{subsec:detect}

Most honeynet operators rightfully fear liability if intruders use the
honeynet as a stepping stone to attack systems of third parties.  Therefore,
two main techniques are deployed to minimize the risk of a successful attack
on third parties' systems: First, outgoing traffic is heavily restricted to
values like only 15 outgoing TCP connections per day. Such extreme means of
traffic shaping are trivial to detect: We only have to initiate a large number
of outgoing connections (SYN packets to other hosts) and watch if after some
number of connections we get blocked by the honeywall. 

The second means of avoiding outbound attacks by honeynets is the usage of a
piece of software named \buzz{snort\_inline} \cite{snortinline:homepage}. This
software acts as an inline packet modification engine rewriting what looks
like something considered dangerous to something harmless. The example for
this technique used by the honeynet project all the time is the replacement of
the string \texttt{/bin/sh} in shellcode with the string
\texttt{/ben/sh\cite{gen2:2003}}.  Another example is the replacement of some
characteristic patterns in network traffic in order to render attacks useless,
as shown in Figure \ref{fig:snort}.

\begin{figure}[htb]
\begin{verbatim}
 alert ip $HONEYNET any -> $EXTERNAL_NET any 
   (msg:"SHELLCODE x86 stealth NOOP"; sid:651; 
    content:"|EB 02 EB 02 EB 02|"; 
    replace:"|24 00 99 DE 6C 3E|";)
\end{verbatim}
\caption{\com{Snort\_inline}-rule to modify packets with x86-shellcode.}
\label{fig:snort}
\end{figure}

This allows us to detect a honeywall trivially by constructing some network
communication containing strings which match \com{snort\_inline}'s rewriting
database and check if the communication is received in unaltered form. 

We are able to implement both ways to detect the honeywall and our experiments
show that a honewall can be easily detected this way.

\subsection{Circumventing Logging on a Honeynet}
\label{sub:circ}

\subsubsection{Overview}
\label{subsub:circ-overview}

Logging on a honeynet can be circumvented by making logged data
unintelligible, making logging itself unreliable or using ways not to be
logged at all.

There is little we can do about the logging of all network communications done
by the honeywall. We might be able to penetrate the honeywall and wipe data
from the machine, but we cannot rely on this option. This means that we have
to use cryptography in network communications whenever possible to avoid
logging. The main problem in doing so is how to deliver our initial exploit to
the target in an encrypted fashion. In many instances we will not be able to
deploy encryption when transmitting the initial exploit. Notable exceptions
are exploits delivered via HTTPS: If the target offers the HTTPS service we
have a wide range of opportunities to exploit the webserver, vulnerable
CGI-Programs or other programs accessible via the web interface while being
able to communicate in encrypted form and therefore unintelligible for network
traffic logging. 

To circumventing logging by Sebek we again take a look at how Sebek obtains
this data: Sebek has all data available which is gained by the \com{sys\_read}
kernel function. The PID and UID of the process calling \com{read()} is
recorded, as well the number of the filedescriptor being read, the first 12
characters of the filename of the program calling \com{read()} and the data
being read itself. Nearly as interesting as what is logged is the observation
what is not logged. For example, neither is the path of the file being
accessed logged nor the path of the program calling \com{read()}. This means
the adversary doing forensic analysis is restricted to guesswork on what is
read by whom. This allows us to clutter up Sebek logging with arbitrary data.
For example, we can create a file looking like a legitimate shadow-password
file in \com{/tmp} and read it via \com{cat}. The adversary will probably at
first glance assume we were able to read the real shadow-file. We can also
copy binaries, which we are going to use, to random names and in doing so give
the adversary a hard time guessing which programs were executed. While we have
no direct control over the PID, we can add frequent executions of \com{fork()}
coupled with termination of the parent to keep the PID being a moving target.
If such deceptive measures are executed in an automated yet well planned
fashion, we can mix the commands we issue to directly accomplish our
objectives with thousands of dummy commands.  Hence, we can generate so much
logging data that forensic analysis will be made much more complicated.

Another option is to make logging by Sebek unreliable. As shown in section
\ref{sub:detect}, this can be archived via the \buzz{dd-attack}, in which
dropping of Sebek packages should take place. The percentage of packages
dropped by the \buzz{dd-attack} varies greatly. So while this method of
circumventing is not fully reliable, it can be expected that the adversary
will not be able to fully reconstruct a session while a \buzz{dd-attack} takes
place.

The preferred method to circumvent logging is avoiding it at all.  In order to
avoid recording of the activities on a honeynet which is using Sebek, we have
to avoid using \com{read()} system calls, since all data which is read via
\com{read()} system calls is transfered over the network to a logging host.
One way of avoiding \com{read()} is to use \com{mmap()} to map files directly
into memory and therefore bypassing logging. This allows us to read all
regular files on the system without Sebek noticing this at all. Unfortunately,
device files and files on special filesystems like \com{/proc} can not be read
in this manner.

For reading data from the network we can exchange the \com{read()} system call
for \com{recv()} which will not be logged by Sebek. Unfortunately, this is of
little practical value since we have to assume that all network traffic is
captured by the adversary so we have to encrypt network traffic. So while
using \com{recv()} provides only limited help to keep data secure at least it
hides the fact from the adversary \emph{which} process is participating in the
network communication. 

It is notable that executing programs via \com{exec()} does not involve
\com{read()}. This basically means that we can in theory execute arbitrary
programs without logging by Sebek taking place. It turns out that this is only
true to a limited extent: At program startup, the dynamic linker loads various
dynamic libraries into memory. This is done mostly via \com{mmap()} but also
sometimes via \com{read()}. So if we execute a program, there is a chance that
Sebek is logging a \com{read()}-call of a library by that program.

\subsubsection{Kebes: A circumvention device}
\label{subsub:circ-kebes}

To experiment with some of the limitations in Sebek, we constructed a
proof-of-concept toolkit called \buzz{Kebes}. Kebes is written in the
high-level programming language \buzz{Python} and is designed to allow a wide
variety of actions being taken on the target system without \com{read()} being
called.

While Kebes uses a layered approach in communication and is therefore not
restricted in the type of communication it uses, we opted for simplicity and
decided to use a traditional client/server setup: The server resides on the
target, listening on a TCP socket and forking childs for handling connecting
clients. 

The TCP connection is used by a layer that we call the \buzz{crypt layer},
which provides an AES-encrypted message-based communication channel for Kebes.
Besides the encryption itself, the crypt layer provides some support services
like random padding and compression via \buzz{zlib} where we use a random
compression level between $0$ and $9$ to further obfuscate the message length.
The encryption is set up by utilizing a Diffie-Hellman key exchange, so there
are no pre-shared secrets between the Kebes server and client. Using plain
Diffie-Hellman makes the crypt layer vulnerable to man-in-the-middle attacks.
But our concern constructing Kebes was not to defeat the active attacker; such
a scheme must ultimately fail since the active attacker probably has also
complete control of the target host we use to run the Kebes server. Rather our
goal was to make ex-post analysis of network, filesystem and Sebek data hard
if not impossible.

Above the crypt layer we use the so called \buzz{Kebes layer}. The Kebes layer
consists of serialized Python data structures: Each command message consists
of a list of commands whereas each command consists of a tag considered by the
server as an opaque identifier of the instance of command invocation designed
to allow asynchronous return of replies. The second element is a string
identifying the command to be executed and the third element is a list of
parameters, which can consist of command specific Python objects.

Tapping in the strengths of the highly dynamic Python language, the Kebes
server initially understands just a single command: \com{ADDCOMMAND}.  All
further commands desired by the client are dynamically loaded via the network
in the server process. This basically means that the server is only a
communication stub which is dynamically extended via an encrypted network
channel. This does not only make maintenance much easier, since updates can be
pushed by the client into the server without restarting the server, but also
should make forensic analysis of Kebes' inner workings much harder: As long as
no paging or core dump of the Kebes process occurs, there should be no
permanent record of the more advanced functionality added to the server by
pushing the code for that functionality to the client.

The Kebes commands implemented this way include besides trivial to implement
commands for listing directories, getting file information, creating files and
getting system information some more complex commands: Deleting files is
implemented in the most secure manner possible in a portable way in a high
level language. We rename the file to a random name, overwrite its content by
a slightly longer random pattern, sync the file to disk and repeat the whole
procedure various times before finally unlinking the file.

Reading a file is implemented by using \com{mmap()} to instruct the virtual
memory system to map the file contents into the address space of the Kebes
server. A copy of this memory area is then simply sent over the network to the
Kebes clients. As shown above, this method of reading files is not logged by
Sebek. 

Files are executed by first redirecting file descriptors for \texttt{stdout}
and \texttt{stderr} to write to files and creating a file from which
\texttt{stdin} will read. While using a file for \texttt{stdin} creates a risk
of being recovered -- even if we try secure deletion as outlined above -- we
see little additional risk of compromising extra information, since data read
by the process we are executing can be logged by Sebek anyway. Redirecting
output to a file is the only way to allow us using \com{mmap()} to read the
output without calling \com{read()}, although we gain this advantage by the
increased risk that secure deletion of this files is unsuccessful and forensic
disk-analysis will reveal the output saved in the files. After the files are
set up, the Sebek server forks, uses \com{exec()} to execute the desired
commands and after termination of the child process returns process id,
status, and output to the Kebes client.

We also created an extended version of the execution command which is able to
receive a binary from the client, saves it under an arbitrary name, executes
it and finally securely deletes the binary again. This version of the
execution command can also create a temporary copy of an existing command on
the target host with an arbitrary name, executes the copy and then deletes it.
While we have only limited control on where the programs executed by us use
\com{read()}, we make analysis of data collected by Sebek much harder by
controlling the process names under which the read activity is logged.

For highly sensitive code, which should fail under no circumstances in the
hands of the adversary, we prototyped a way to execute code without ever
writing it to disk. To support this, the Kebes server can compile an extension
module based on \texttt{C}-Code for Python called \buzz{shellcode-executer} on
the fly. The shellcode-executer extension takes a Python string -- which can
contain $0$-bytes -- and lets the processor directly jump into that string
allowing the execution of arbitrary machine code by Kebes. This idea could be
extended until Kebes could execute complete ELF binaries received by the
network as pioneered in \cite{m1lt0n04:_keepin_safe}.

The main challenge when implementing Kebes was to get the entropy for the
secret in the initial Diffie-Hellman key exchange: Since \texttt{/dev/random}
can not be read by \com{mmap()}, we would have to resort to \com{read()},
which would allow an adversary to gain knowledge of the entropy used to set up
the Diffie-Hellman key exchange and thus reconstruct the communication keys.
We first contemplated to \com{fork()} another process, which would perform an
\buzz{dd-attack} by pushing as much data as possible through \com{read()}.
This would be done by reading \texttt{/dev/zero} hoping that this would result
in an overload of the network carrying the data captured by Sebek and thus
result in not all data processed by \com{read()} being logged. While it turned
out that this generally works as expected, it is extremely unstealthy. We now
use clock jitter, thread races and the contents of some files to get the
entropy from. 

We start two threads doing exactly the same: for every file we find in the
\texttt{/var} and \texttt{/tmp} filesystem of the host, we add to a temporary
list:

\begin{itemize}

	\item A high resolution timestamp when processing started.

	\item The ID of the thread to encode the hard to predict fact which
		thread is processing a file first.

	\item SHA-1 of the file in question if the file is readable and
		smaller than one megabyte; if the file is readable but bigger,
		the last megabyte of the file. Reading is done via
		\com{mmap()} to avoid logging by Sebek.

	\item A high resolution timestamp when processing stopped.

\end{itemize}

Every item from this temporary list is then stirred together with it's list
position and another timestamp in the $1960$ Bit entropy pool. While this
method is probably unsuitable for strong cryptography, it seems good enough to
seriously slow down forensic examination of network traffic.

\subsubsection{Testing the NoSEBrEaK Ideas}
\label{subsub:cirv-test}

To test the effectiveness in defeating Sebek by our \mbox{NoSEBrEaK} ideas, we
prepared a Linux-honeynet running Sebek 2.1.7. This machine contains a
CGI-Program which we prepared to contain a stack-overflow. We also planted a
SUID-root executable with an stack-overflow in parameter processing. Instead
of this vulnerabilities introduced by ourselves, we could have exploited known
vulnerabilities in the Linux installation on the honeynet, but for testing
purposes we wanted to be somewhat independent of distribution flavors and
software versions.

We prepared an exploit for the CGI-Program utilizing shellcode that is
able to load a binary from the network and execute it. As the binary
to be transferred, we created a compound binary consisting of the Python
runtime environment the byte code compiled Kebes server. We also created an
exploit for the vulnerable SUID-root tool, which is able to execute
arbitrary programs with \com{exec()}. Finally, we created a Python
program packaged again as an compound binary which accesses kernel internal
data structures to find the Sebek module, guesses the values used for
network addresses, ports and magic value in the Sebek module and
the addresses of the module's \com{init()} and \com{cleanup()}
functions. This program is accompanied by a minimal kernel-module
which just enables the user to jump to arbitrarily locations in the kernel.
By calling this module with the address of Sebek's \com{cleanup()}
function we can force Sebek to disable itself.

Equipped with this toolset, we first connect via HTTPS to the webserver on our
target. Using HTTPS practically renders logging of network traffic ineffective
-- although analysis of traffic patterns and source address is still possible.
We deliver the shellcode via a stack-overflow in the path and the compound
binary in the request body. The binary is read by the shellcode, written to
disk and executed. This starts the Kebes server which binds to a TCP ports and
listens to incoming connections while running under user ID \texttt{www-data},
like the webserver it was spawned off is doing. We connect with our Kebes
client to the Kebes server and add dynamically the needed commands to the
server. Then we write the binaries used to deactivate Sebek to disk and start
the vulnerable SUID-root program with parameters exploiting the vulnerability
and starting the Sebek detection and removal program as root. Finally, we
again exploit the SUID-root binary, this time to start a Kebes server with
root permissions.

After this few relatively easy steps we earned full control over the honeynet
machine while none of our activities were logged by Sebek and the recording of
network traffic only left the adversary with some encrypted traffic and the
clear-text Diffie-Hellman key exchange.

Therefore our experiment rebuts our original assumption that even if an
attacker manages to disable Sebek we would at least gain a transcript of the
methods used to disable Sebek. We can not trust any data that is logged on a
Sebek-based honeynet if there has some encrypted communication between the
honeynet an a possible attacker taken place.

\section{Further work}
\label{sec:further}

There is a broad range of possibilities to enhance our work in further
attacking Sebek. But also several approaches in hardening Sebek against
attacks come to mind.

For further attacking Sebek we see lots of possibilities. Some of them
include: 

\begin{itemize}

	\item Sebek leaves many traces in memory: It is a kernel module and
		thus uses some special data structures, it uses characteristic
		variables and builds UDP packets in memory. Therefore it
		should be possible to search through memory and detect its
		existence in several other places than by looking for module
		structures.

	\item Disable Sebek by patching the Sebek kernel module at runtime to
		skip the call to the logging function.
        
	\item Burneye, an executable encryption program which is suited to
		protect ELF binaries on the Intel x86 Linux operating system
		\cite{burneye:homepage}, is very popular among blackhats. With
		the help of Sebek, the password used by an intruder to decrypt
		the binary can easily be observed \cite{Sebek:2003}. The
		implementation of a challenge-response technique for burneye
		in order to protect the binary would make the decryption much
		harder.

	\item Hook into the low level networking layer at the level of the
		network drivers to unhide Sebek's logging packages.

\end{itemize}

If we turn the table again and take the view of a honeynet researcher, we see
some topics of interest in order to harden Sebek and its general approach to
log activities of intruders:

\begin{itemize}

	\item Obviously, transforming Sebek to be a kernel patch instead of a
		loadable kernel module should be aspired. Detecting and
		removing Sebek if it is not dynamically loaded as a kernel
		module would be much more difficult and installation after a
		reboot would become a non-issue.  On the other hand,
		deployment of Sebek would get complicated by doing so. In our
		opinion this is actually desirable, since honeynets are
		dangerous and ethically problematic tools and entities wishing
		to deploy honeypots should be willing and able to got to the
		pains of patching and compiling a kernel.

	\item Adding further cleaning of module structures in memory after
		installing the Sebek kernel module, using a less predictable
		way of generating the ``random'' symbol and module names.
  
	\item Further obfuscating the placement of Sebek's variables in memory
		and adding decoy variables.

	\item Deploying polymorphic techniques to further obfuscate Sebek in
		memory.                       

	\item Using source and destination ports and MAC addressees in
		addition to the magic number when identifying Sebek packages
		to extend search space from $2^{48}$ to $2^{160}$ Packets when
		brute-forcing.

\end{itemize}

\section{Conclusions}
\label{sec:con}

We have shown that Sebek can be detected, disabled and it's logging
circumvented. This knowledge lets us take a completely different view on data
obtained by honeynets and on data not obtained by honeynets. While
unsophisticated attackers might not be able to circumvent honeynets or not
even try to do so, we assume that sophisticated attackers can detect honeynets
and disable logging on them if this fits their objectives. If there are
already very advanced techniques of detecting and disabling honeynets
discussed in the open wild
\cite{corey03:_local_honey_ident,corey04:_advan_honey_pot_ident}, we have to
assume that there are much more evolved techniques available to highly
advanced attackers. This underlines our apprehension that honeynet technology
is only able to gather information of the common crowd of unsophisticated
attackers, but has a very small probability of gathering significant
information on advanced attackers which would be of much more value to
researchers. 

\section{Acknowledgements}
\label{sec:ack}

We owe thanks to Felix G\"artner for making our research possible and Phillip
Maih\"ofer for providing us with test platforms.

Thorsten Holz was supported by the Deutsche Forschungsgemeinschaft (DFG) as a
research student in the DFG-Graduiertenkolleg ``Software für mobile
Kommunikationssysteme'' at \mbox{RWTH} Aachen University.

\bibliographystyle{ieeetr}
\bibliography{NoSEBrEaK}
\end{document}